\documentclass[12pt]{article}

\usepackage{sbc-template}
\usepackage{graphicx,url}
\usepackage[utf8]{inputenc}

\usepackage{multirow}
\usepackage{amssymb}
\usepackage[table,xcdraw]{xcolor}
\usepackage{pgfplots}
\usepackage{multirow, booktabs, makecell}

\usepackage{lipsum}  

\title{\Large{VNF-Cache: An In-Network Key-Value Store Cache Based on Network Function Virtualization}}

\author{
Bruno E. Farias,
José Flauzino,
Elias P. Duarte Jr.
}

\address{
Federal University of Paraná (UFPR), Dept. of Informatics, \\ Curitiba, PR, Brasil
  \email{\{bef18,jwvflauzino,elias\}@inf.ufpr.br}
}

\begin{document} 
\pagestyle{plain}
\maketitle
\thispagestyle{plain}

\begin{abstract}

With the exponential growth of the amount of data available on the Internet, optimizing the response time and resource usage for data access becomes essential. Caches are an effective solution that brings data closer to clients, eliminating repetitive requests to servers. This paper presents VNF-Cache, a caching service for geographically remote key-value databases. VNF-Cache is an NFV-COIN (Network Function Virtualization-Computing In The Network) service, a technology undergoing standardization by the IETF that enables the implementation of arbitrary services directly in the network. VNF-Cache intercepts network packets, processes, stores, and sends values directly to clients when possible. Through a proof-of-concept implementation and experiments conducted with geographically dispersed servers in Brazil, the United States, and Japan, significant reductions in response time and increases in the number of requests processed per second were observed.


\end{abstract}

\section{Introduction}
\label{sec:intro}

Caches are a fundamental mechanism for improving the efficiency of data access over the Internet. This paper introduces VNF-Cache, a caching service for key-value databases built with Network Functions Virtualization (NFV) technology. NFV enables the implementation of network functions as software based on virtualization technologies \cite{NFV_WHITE_PAPER}. Given the wide variety of network functions that can be deployed using NFV, the NFV-MANO (NFV Management and Orchestration) reference architecture was established to standardize their management and ensure interoperability across NFV ecosystems \cite{NFV_MANO_man}. This architecture defines the lifecycle management of Virtualized Network Functions (VNFs) and the computational resources they require \cite{NFV_soa_research}.

Recently, the NFV-COIN (NFV - COmputing In the Network) architecture has been proposed to support the deployment of arbitrary and innovative services directly within the network, through the COIN (COmputing In-the-Network) paradigm \cite{venancio2022nfv}. NFV-COIN, is currently undergoing standardization within the IETF, which has already published two drafts defining two foundation components: one outlining the problem statement \cite{nfv-coin-problemstatement} and another specifying the architectural framework \cite{nfv-coin-framework}.

The VNF-Cache proposed in this work is designed as an NFV-COIN service. Its purpose is to optimize client access to geographically distant key-value databases by placing data closer to end users. By reducing the distance that requests and responses must traverse, VNF-Cache lowers access latency and increases the throughput of processed requests. Additionally, by avoiding unnecessary end-to-end communication with the backend server, the service contributes to a more efficient utilization of network resources.

To provide key-value caching capabilities, VNF-Cache must be located at an intermediate point in the network between clients and the corresponding database servers. Its operation begins by identifying and filtering relevant traffic, after which it processes and stores the values associated with keys requested by clients. When cached data is available, VNF-Cache responds directly to the client; otherwise, the request is forwarded to the server, and the corresponding response is intercepted and stored, when applicable, to serve future requests.
 
The VNF-Cache architecture is described in detail, along with a proof-of-concept prototype. An empirical evaluation was conducted across three experimental scenarios: (A) in the first scenario a client, VNF-Cache, and key-value database server are all located in close proximity; (B) in the second scenario, the client and VNF-Cache are close to each other, while the server is geographically distant; and finally (C) in the third scenario the client, VNF-Cache, and server all located distant to each other. The remote instances of VNF-Cache and the database servers were deployed on Amazon Elastic Compute Cloud (Amazon EC2), a cloud computing service from Amazon Web Services (AWS)\footnote{https://aws.amazon.com/pt/}. Virtual machines were instantiated in Curitiba and São Paulo (Brazil), Ohio (USA), and Tokyo (Japan). The results show that, in scenarios where clients and servers are geographically distant, introducing VNF-Cache can substantially reduce request response times and increase the throughput of processed requests.

The remainder of this work is organized as follows. Section \ref{sec-related} presents an overview of NFV and NFV-COIN technologies, as well as related work. Section \ref{sec-vnfCache} describes the VNF-Cache, detailing its architecture and the prototype that was implemented. Section \ref{sec-evaluation} reports the experimental results. Finally, Section \ref{sec-conclusion} provides concluding remarks and outlines directions for future work. 
\section{NFV, NFV-COIN \& Related Work}
\label{sec-related}

Network Functions Virtualization, or NFV, is a technology for implementing network services using virtualization, such as Virtual Machines (VMs) and containers, which can run on off-the-shelf hardware. A wide variety of network services can be implemented with NFV, such as routers, Virtual Private Networks (VPNs), traffic analyzers, firewalls, Content Delivery Networks (CDNs), among others \cite{NFV_WHITE_PAPER}. Before NFV technology became available, the single alternative was to have those services implemented in specialized hardware, which is a substantially less flexible approach. NFV has changed the global communication networks scenario, by broadening the scope of vendors of network funtions and services, which are made available through global marketplaces \cite{bondan2019fende}.

In order to establish standards and develop an architecture for managing the life cycle of virtualized network functions and services is the MANagement and Orchestration (MANO) model \cite{NFV_MANO_man}. MANO emerged as an architecture to enable interoperability between NFV systems. To this end, the architecture encompasses not only the lifecycle management of VNFs and but also of the computational resources they use. A very large number of NFV platforms based on MANO are currently available \cite{tacker2025tacker,etsiosm2025open,ger_orq_nfv}. Note that a key component of an NFV platform is the VNF execution platform \cite{garcia2019design,garcia2020design}, which is responsible for the actual execution of the network functions \cite{huff2018holistic,tavares2018niep}. 

The NFV-MANO architecture is basically composed of three fundamental and interdependent blocks: the Virtualized Infrastructure Manager (VIM), the VNF Manager (VNFM), and the NFV Orchestrator (NFVO), described next.
The VIMs are responsible for controlling and managing the physical infrastructure and resources available for virtualization. A single VIM can be responsible for managing only one type of computing resource at a time (such as processing, memory, or network), or several types of resources simultaneously. It must be able to efficiently distribute the resources made available to an NFVI (or NFV Infrastructure) in order to meet the needs of each VNF without impacting the execution of the others. The NFVI is the set of physical hardware together with the virtualization layer that runs the VNFs.

The VNF Manager (VNFM) is responsible for managing the lifecycle of VNF instances; that is, this module is responsible for instantiating and removing instances, performing configurations and updates, as well as monitoring and scaling functions \cite{venancio2021beyond}. A single VNFM is responsible for managing several virtualized network functions of different types, the EM (Element Management) instrumentalizes the VNF for the purpose of management and interactions with the VNFM \cite{fulber2023etsi}.

The NFV Orchestrator (NFVO) has two main functions: managing the NFVI across multiple existing VIMs and coordinating the lifecycle of network services. The NFVO is the component that enables the composition of VNFs  that for a Service Function Chain (SFC) to create complex services \cite{Halpern-2015}. It is possible to combine multiple different VNFs, each solving some specific problem, in different topologies \cite{fulber2020network}. Besides SFC composition itself \cite{fulber2020cusco,garcia2020nsh}, SFC deployment is a key area: how to properly allocate an SFC across a single or multiple datacenters \cite{fulber2023customizable}
The Multi-SFC is an approach to allow the instatiation of SFCs across multiple clouds, autonomous systems, and orchestrators \cite{fulber2024breaking,huff2020building}.

In summary, virtualized network functions are managed by VNFMs and run on Network Function Virtualization Infrastructures (NFVIs). These, in turn, are managed by VIMs, which are integrated with and managed by the NFVO. The NFVO is responsible for distributing all computing resources among the VIMs and managing other necessary network services. All these elements are generally made available through what are called NFV Platforms: \cite{tacker2025tacker, etsiosm2025open, ger_orq_nfv}. 

As previously presented, NFV technology can enable the implementation of a variety of network functions, from the simplest to the most complex services. The NFV-COIN \cite{venancio2022nfv} architecture expands this scope to allow arbitrary and innovative services directly within the network, through the \emph{COmputing In the Network} (COIN) paradigm. Being software-based and virtualized, NFV-COIN technology offers great flexibility for deploying new native network features in a cloud-edge-core continuum. NFV-COIN is on the IETF standardization track, with Drafts published for both the problem statement \cite{nfv-coin-problemstatement}, and interface to network functions \cite{nfv-coin-framework}. Several NFV-COIN services have already been proposed, such as failure detectors \cite{turchetti2015implementation,turchetti2017nfv}, consensus \cite{venancio2021vnf}, and reliable and ordered message broadcasting \cite{venancio2019nfv}.

Despite the great advances of NFV technology, it will be only mass adopted when it guarantees the availability levels of the telecommunications industry. Advances have been made from this point of view too, with strategies to improve both fault \cite{venancio2022nham,venancio2024highly} and intrusion-tolerance \cite{venancio2024dependable}.

\subsection*{Related Work: NFV \& Caches} In recent years, several works have been developed involving the union of NFV and caches. For example, \cite{zhuang2019sdn}  discusses the possibility of applying NFV-based caches to minimize content retrieval time in Internet-of-Vehicles (IoV) systems. According to the authors, the use of this type of cache can facilitate the deployment of services and the dissemination of their content, enabling better reliability and efficiency of IoV services.

Earlier, \cite{clayman2018virtualized} had proposed an architecture for video streaming called \emph{Server and Network Assisted Dynamic Adaptive Streaming over HTTP} (SAND). In this architecture, virtualized cache instances are created according to content demand. Furthermore, the authors also discuss the positioning of these instances in the network graph, based on characteristics such as path bandwidth, location, and the number of network clients.

Another relevant related work is \cite{liu2017sdn}, which discusses the remarkable performance gains that the application of NFV caches can bring. In addition to the flexibility, dynamism, and scalability enabled by the use of NFV, the authors also highlight the possibility of offering caching services to service providers and network operators using the same infrastructure.

\section{An In-Network Cache Service for Key-Value Databases}
\label{sec-vnfCache}

This section describes the VNF-Cache in detail. In terms of functionality, the VNF-Cache is a caching service designed for Key-Value Store (KVS) databases. This is a type of non-relational database that persists data by associating a unique key with each stored data item \cite{KVS_practical_overview}. Using this type of database allows the application developer to store data without using schemas, i.e., without the traditional relational method of having predefined rows and columns. This provides greater flexibility in database design, as well as an improved quality of the corresponding programming code.

\begin{figure}[!htb]
    \centering
    \begin{minipage}{.57\textwidth}
        \centering
        \includegraphics[width=\linewidth]{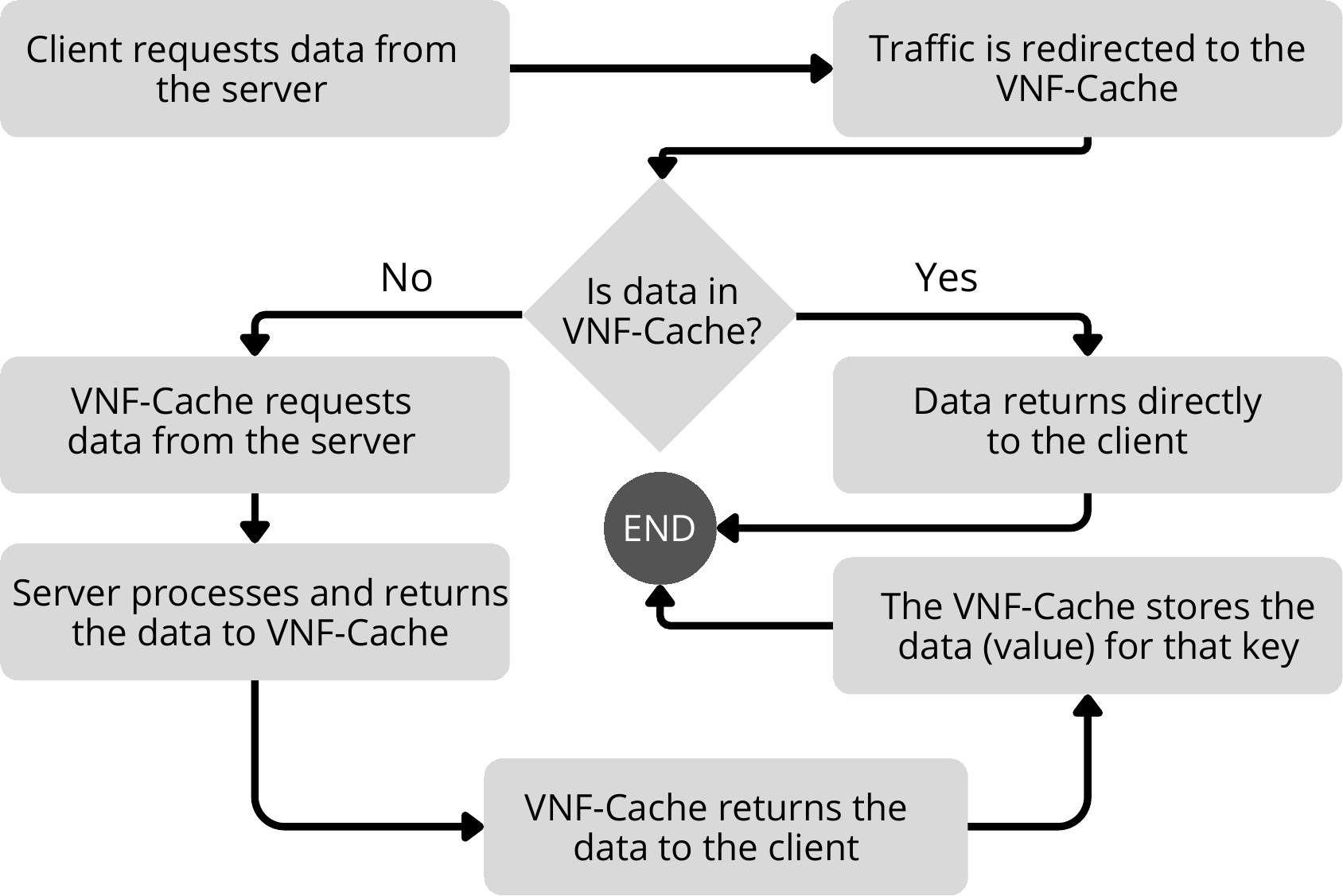}
        \caption{VNF-Cache operation.}
        \label{fig:vnfcache_fluxogram}
    \end{minipage}
    \begin{minipage}{.423\textwidth}
        \centering
\includegraphics[width=.9\linewidth]{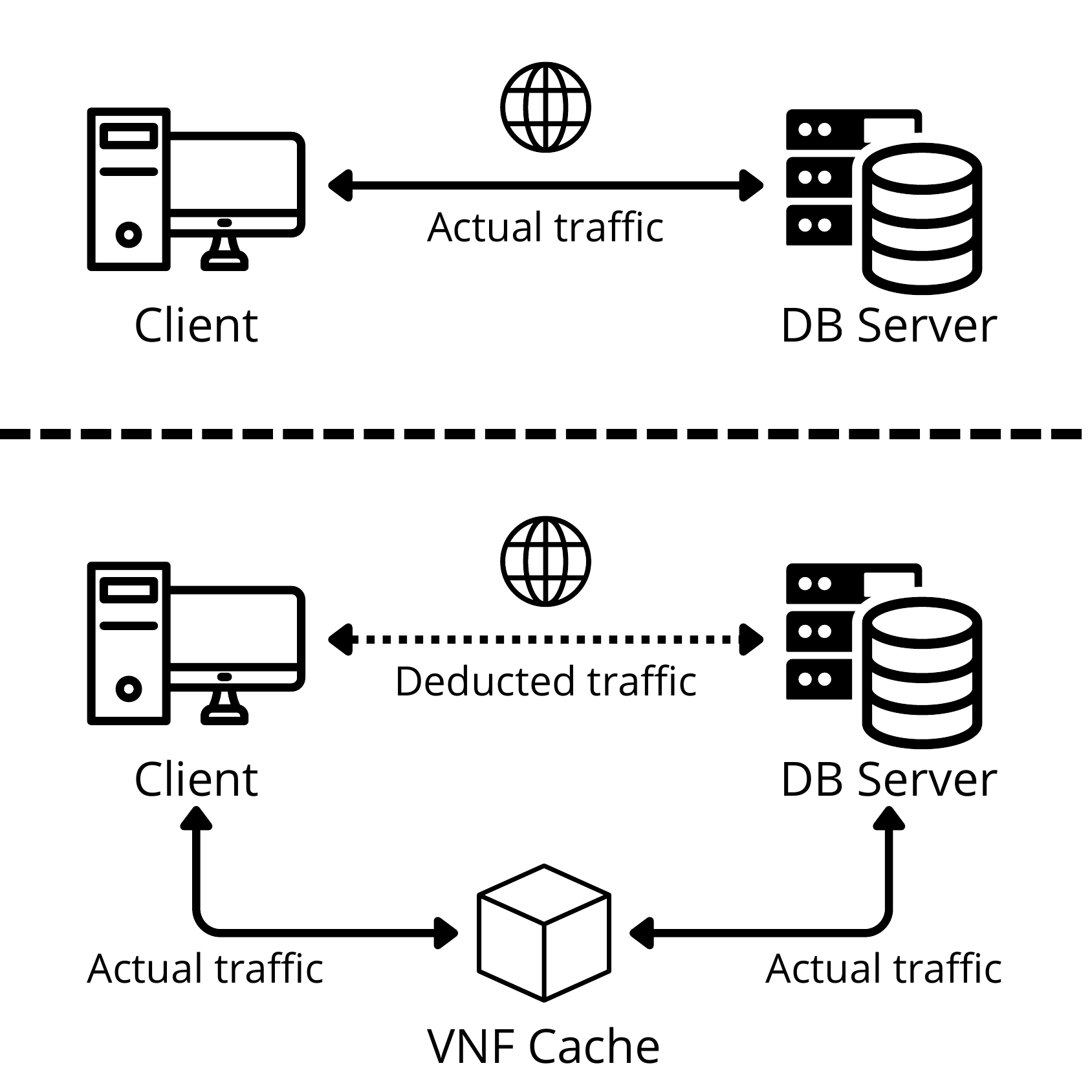}
        \caption{Example of network architecture with and without VNF-Cache.}
        \label{fig:vnfcache_network_example}
    \end{minipage}
\end{figure}

Overall, the main objective of the VNF-Cache is to bring data closer to a client. However, one of the differences from other caching services is that it stores data directly in the network, along the path between the client and the remote server. The VNF-Cache intercepts and processes the database packets and stores the values of keys requested by clients. On processing a client request, if the key is valid in the cache, the corresponding value is returned directly to the client, eliminating the need to forward packets to the server. Figure \ref{fig:vnfcache_fluxogram} illustrates the sequence of steps of the VNF-Cache operation.

Figure \ref{fig:vnfcache_network_example} shows the network with and without a VNF-Cache. In the figure, the deduced traffic is defined as the traffic that the client deduces it is generating on the network, which may differ from the actual traffic that is being carried on the network. 

\subsection{VNF-Cache Architecture}
\label{subsec_vnfCache_arch}

\begin{figure}[!htb]
\centering
\includegraphics[width=12cm]{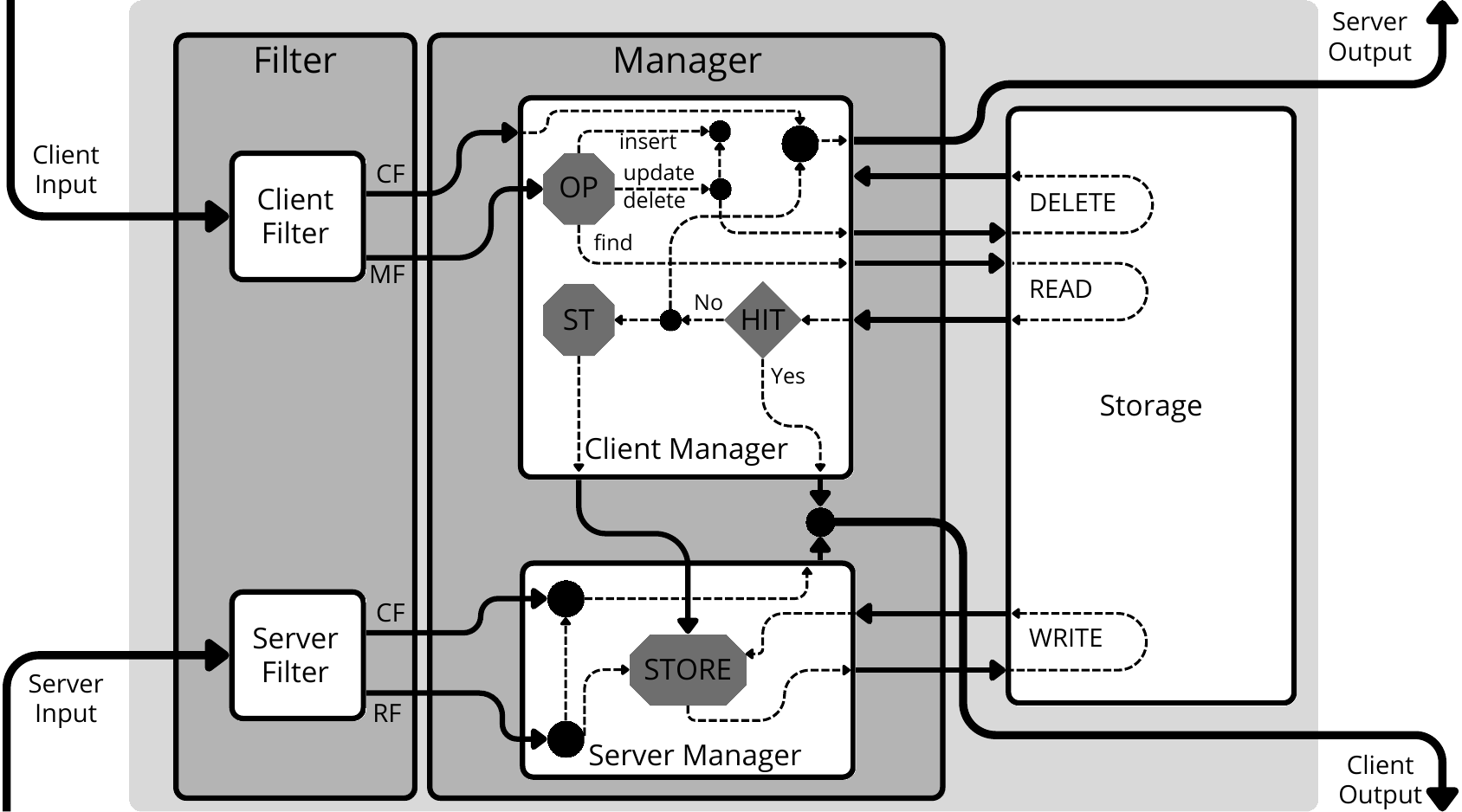}
\caption{VNF-Cache architecture.}
\label{fig:vnfcache_architecture}
\end{figure}

The VNF-Cache architecture consists of three modules: \emph{Filter}, a \emph{Manager}, and \emph{Storage}, illustrated in Figure \ref{fig:vnfcache_architecture} and described below. The VNF-Cache \emph{Filter}, or simply \emph{Filter}, is the module responsible for filtering packets received by the VNF-Cache, whether sent by clients (\emph{Client Filter}) or the server (\emph{Server Filter}). These two submodules receive and filter network packets into three possible flows: the \emph{Manipulation Flow} (MF), the \emph{Response Flow} (RF), and the \emph{Coordination Flow} (CF), described next.

The data manipulation flow (MF) consists of packets sent by clients containing data manipulation operations, such as searches (queries), insertions, updates, and deletions. The response flow (RF) consists of packets sent by servers to respond to the requests sent by clients in the MF flow. Finally, the coordination flow (CF) consists of other packets, i.e., those sent by clients or servers with some management purpose, such as maintaining the connection between clients and servers or monitoring server availability.

The \emph{Client Filter} is responsible for filtering packets originating from the client, separating them into the MF flow and the CF flow. As mentioned above, the MF flow consists of the packets containing data manipulation operations issued by clients to the database server. The CF flow consists of packets issued to perform basic management tasks between client and server, in this case, sent from the client to the server. Similarly, the \emph{Server Filter} filters packets coming from the server to a client, separating them into the RF flow (which consists of the packets produced to respond to requests made by clients in the MF flow) and the CF flow (which, in this case, is from the server to the client).

The \emph{Manager} is the main management module of the VNF-Cache. Similar to the \emph{Filter}, the \emph{Manager} is also composed of two submodules: the \emph{Client Manager} and the \emph{Server Manager}. The \emph{Client Manager} receives the two packet flows from the \emph{Client Filter} and processes them as needed: the MF flow is handled directly by the VNF-Cache storage module \emph{Storage} (which is presented below), performing data reads, updates, and deletions according to the operations. The CF flow is forwarded directly to the server. Similarly, the \emph{Server Manager} receives the two packet flows from the \emph{Server Filter}. However, all packets of both flows are forwarded from the \emph{Server Manager}  to the client. The difference in how the \emph{Server Manager} processes both flows is as follows. While the CF flow is immediately forwarded to clients, the RF flow packets undergo extra processing, intending to store the data returned by the server in the VNF-Cache \emph{Storage}.

Finally, the \emph{Storage} module is responsible for storing the keys and their respective values at the VNF-Cache. This module has two main functionalities: (\textit{i}) returning the value of a key requested by the \emph{Client Manager} and (\textit{ii}) storing the value of a key captured by the \emph{Server Manager} and requested by the \emph{Client Manager}.

The two main flows in the VNF-Cache are generated after a client requests a key. When a client requests a key, the \emph{Client Filter} filters this packet in the data handling flow (MF). Then, the \emph{Client Manager} determines the operation type, which for a data query is \emph{find}, and makes a request to the \textit{Storage} for the value of that key. The storage module either returns a cache miss or a cache hit. If the request results in a cache hit, the \emph{Client Manager} returns the packet directly to the client. On the other hand, if the request is a cache miss, the \emph{Client Manager} forwards the packet to the server and informs the \emph{Server Manager} that the requested key is not in the cache and that, if possible, the value that will be returned from the server should be stored (in the \textit{Storage} module).

\subsection{VNF-Cache Implementation}
\label{subsec_vnfCache_impl}

VNF-Cache was implemented in Python\footnote{https://www.python.org}. Multiple Python libraries were employed, such as Scapy and PyShark. The database was MongoDB\footnote{https://www.mongodb.com}, a JavaScript Object Notation (JSON) document-oriented database. Although MongoDB is not exclusively a key-value database, it can be used as such by storing data in the form of flexible documents. Each data item is linked to a unique, automatically generated ``index", which works as the key in the key-value context. It is important to note that VNF-Cache can be easily adapted for use with other key-value databases, such as Redis, for example. To enable communication between clients and MongoDB database servers, the  PyMongo\footnote{https://pymongo.readthedocs.io/en/stable/index.html} library was used.

To enable the communication between clients and the MongoDB database server, the PyMongo library uses network packets based on the IP network-layer protocol and the TCP transport protocol. The VNF-Cache filter only captures packets with operation identifier 2013, which is the standard type for PyMongo search, insert, delete, and update requests. PyMongo packets have a standard 25-byte header, separated into 7 fields, described next.

The first field (4 bytes) is the \emph{length}, which contains the total size of the TCP segment. The second field (4 bytes) is the \emph{request\_id}, which represents a unique identifier for the request (the same identifier must be present in the response packet). The third field (4 bytes) is the \emph{response\_to}, which contains the response identifier that must be the same as that of a corresponding request. The fourth field (4 bytes) is the \emph{op\_code}, which contains the operation identification number (VNF-Cache only processes packets with the operation code 2013 -- manipulation flow). The fifth field (4 bytes) contains some \emph{flags} for communication between PyMongo and MongoDB. The sixth field (1 byte) is the \emph{payload\_type} and contains the content type of the PyMongo package. Finally, the seventh field (4 bytes) is the \emph{payload\_size}, which contains the total size of the \emph{Binary JSON} (BSON) document contained in that package.

By checking the \emph{payload\_size} and the rest of the payload of the packet, it is possible to extract the complete JSON document sent with PyMongo using the \emph{RawBSONDocument} class from the \emph{bson} library. This enables the VNF-Cache to properly handle packets by capturing and filtering those that are of interest. By using fields of the PyMongo packet header, the VNF-Cache can make the correct decision for each packet. For example, forwarding the packet to the server and storing the response, returning the stored value directly to the client, or applying padding and replacement policies, among others.

In the VNF-Cache implementation, the cache itself is stored as a single Python file. Its operation is exactly as mentioned previously; that is, traffic consisting of network packets destined for the database server is diverted to a specific port of the VNF-Cache. The VNF-Cache then parses the packets and performs the necessary actions required. The diversion of network packets was implemented in a high-performance router and was only possible due to a recent proposal for packet classification through routing directly in the network control plane \cite{flauzino2024anycast}.

To monitor packets received by the cache, a standard Python socket is opened on the specified port and waits for connection requests from clients. When this socket receives a connection request from a client, a new thread is created. In each opened thread, a new socket is created to establish direct communication between the VNF-Cache and the MongoDB server. After the connections between client and cache, and cache and server, are established, the VNF-Cache waits for packets sent by the client.

When a packet is received, the \emph{Filter} module parses the header and separates packets with \emph{op\_code} 2013 and those containing search, insert, update, and delete operations. Other packets are forwarded directly to the MongoDB server or to the client, such as PyMongo connection establishment packets, statistics packets, and monitoring packets.

After capturing a packet with the operation code 2013 (manipulation flow), the VNF-Cache identifies the type of operation and on which data the operation will be performed. It is necessary to determine the method being used, which key is being requested, and, if it is the first request for a key, what value is returned by the server for that key. In order to do this, it is necessary to reconstruct the JSON document sent by PyMongo.

The \emph{Manager} module can then process packets with the following operations: data search, insertion, modification, or deletion. Data search operations employ the term \emph{find} as a keyword. Analogously, insertion, update, and deletion operations employ the keywords \emph{insert}, \emph{update}, and \emph{delete}, respectively.

Finally, the key of the data item of interest is in the \emph{filter} field of the JSON document. According to the MongoDB documentation, the \emph{filter} field allows different combinations of operations, such as searching for keys that are equal to, greater than, or less than the given key, among many others. To simplify this implementation, the focus was only on operations with unique keys. Therefore, in this VNF-Cache implementation, the request key is in the \emph{\$eq} field, which is located in the \emph{filter} field. Optionally, the \emph{\$eq} field can be omitted if it can be extracted from the context. Data insertion operations are not handled by the VNF-Cache, as they do not directly influence what is already stored.

When it is processing a search/read packet for a specific key, the VNF-Cache first checks if this key-value pair is already stored locally. In the VNF-Cache implementation, the data storage module, \emph{Storage}, is implemented as a Python dictionary, using the same key-value pairs as MongoDB. If the requested key is not in the dictionary, the \emph{Client Manager} simply forwards the packet to the MongoDB server and waits for the response packet, storing the value when the response returns and is captured. On the other hand, if the requested key is in the dictionary, the \emph{Client Manager} reconstructs the data packet and forwards it directly back to the client. In this way, the original request packet sent by the client is dropped, never sent to the server.

The implementation of the VNF-Cache was based on the \textit{Write-Invalidate} caching policy  \cite{memory_systems}. Therefore, operations that modify and delete data cause the data to be removed from \emph{Storage}, thus invalidating local data. Thus, when a packet that updates the data of a given key is received, the corresponding key-value pair is removed from the cache and the packet is forwarded to the server, and later the response is properly registered.

Since multiple threads can be running simultaneously and requesting reads and/or writes to the cache dictionary, there is a possibility that two or more threads could modify the same data concurrently, potentially causing inconsistencies. To avoid this problem, the \textit{acquire()} and \textit{release()} primitives are employed, which lock access to the dictionary exclusively for a single thread at a time. These primitives, as well as the instantiation of threads, are from the standard \textit{threading} Python library.

The prototype also provides multiple command-line options. This includes options to define the level of detail of the log, the maximum number of data items the VNF-Cache can hold, as well as the generation of statistical data (e.g., the number of cache hits and misses).
\section{Evaluation}
\label{sec-evaluation}

In this section, we describe experiments conducted with the VNF-Cache across multiple deployment scenarios, varying both its storage capacity and placement relative to the client and server. The evaluation focused on two key metrics: \textit{i}) request response time -- measured as the interval between the instant of time a client issues a request and the instant that the corresponding server response is received by that client; and \textit{ii}) the throughput of processed requests, i.e., the number of requests concluded successfully per time unit. Three scenarios were considered: (A) in the first scenario the client, VNF-Cache, and database server are all located close to each other; (B) in the second scenario the client and VNF-Cache are close to each other, while the server is geographically distant; and finally (C) in the third scenario the client, VNF-Cache, and server all mutually distant. These scenarios enable an assessment of the VNF-Cache efficiency under different network distances between client, cache, and server.

The environment employed to perform the experiments included a single physical machine, along with several combinations of virtual machines, as detailed below. The physical machine is based on an Intel(R) Core(TM) i5-7400 @ 3.0 GHz × 4 processor, with 16 GB of RAM, a 100 Mb/s network interface, and running Ubuntu 20.04.6. That physical host was used to orchestrate the experiments and to host a subset of the virtual machines. Additional virtual machines were instantiated both locally using Kernel-based Virtual Machine (KVM) and remotely via Amazon Elastic Compute Cloud\footnote{https://aws.amazon.com/pt/ec2/?nc2=h\_ql\_prod\_cp\_ec2}, a cloud service from Amazon Web Services (AWS) that enables deploying virtual machines in multiple geographic regions.

MongoDB was used as the key-value store in the experiments, in particular, the randomPhrases database, which stores random text entries. In each experiment, a client sends 30 batches of 1,000 requests for random integer keys, distributed uniformly across the range from 1 to 100. During the experiments, both the response time of each request and the number of requests processed per second were measured.

The VNF-Cache was executed on virtual machines provisioned with two different hardware configurations. On the local physical machine, the virtual machines ran on Ubuntu 20.04 with a virtualized 3 GHz × 2 processor, with 2 GB of RAM and 15 GB of disk storage. On AWS, the virtual machines also ran Ubuntu 20.04, but used a virtualized 2.5 GHz × 1 processor, 1 GB of RAM, and 8 GB of disk storage. The MongoDB database server was hosted on Ubuntu Server 20.04 virtual machines with a 1 GHz processor (in the KVM environment) or a 2.5 GHz processor (on AWS), 1 GB of RAM, and 10 GB of disk storage.

\subsection{Scenario A: Client, VNF-Cache, and Server in Close Proximity}

The first experiment was conducted with the client, VNF-Cache, and the database server, all located close to each other. For this setup, three virtual machines were instantiated on the same physical host. Network redirection policies were configured in the virtual router responsible for inter-VM communication to ensure that all client packets destined for the database server were intercepted and forwarded to the VNF-Cache port. Table \ref{tab:overhead_req_mesma_chave_cen_A} presents the response times of requests issued to the same key in scenarios with and without VNF-Cache.

\begin{table}[!htp]
\centering
\caption{Response times (in milliseconds) for the same key and overhead caused by VNF-Cache in scenario A.}
\label{tab:overhead_req_mesma_chave_cen_A}
\begin{tabular}{| l | r | r | r | r | r | r | r |}
\hline
& \multicolumn{7}{c|}{Order of Requests for a Specific Key}\\ \cline{2 - 8} 
VNF-Cache & 1ª & 2ª & 3ª & 4ª & 5ª & 6ª & {...}\\ 
\hline
No VNF-Cache & 1,12 ms & 1,56 ms & 2,04 ms & 1,26 ms & 1,17 ms & 1,35 ms & {...}\\ 
With VNF-Cache & 10,54 ms & 5,40 ms & 3,87 ms & 3,32 ms & 4,05 ms & 3,93 ms & {...}\\ 
\hline
\textbf{Overhead} & 9,42 ms & 3,84 ms & 1,83 ms & 2,06 ms & 2,88 ms & 2,58 ms & {...}\\ 
\hline
\end{tabular} 
\end{table}

The table shows that, without using the VNF-Cache, the average response time for each request (considering the first 6 requests) ranged from 1.12 to 2.04 milliseconds (ms). The average time was 1.41 ms, with a rate of 535 requests per second. VNF-Cache was then enabled with a maximum capacity of 100 key-value pairs (i.e., the entire key space of the database). This time, considering the cache population, the average response time was 5.18 ms, which is approximately 3.5 times longer. The average throughput decreased to 191 requests per second, corresponding to a reduction of nearly 65\%.

These results indicate that, in scenarios where the client and server are geographically close to each other, the VNF-Cache does not achieve its intended goal of reducing the response time. In such cases, the cache introduces additional processing into the data path: packets that would otherwise flow directly between client and server must instead be intercepted and processed by the VNF-Cache before reaching their destinations. Because the baseline latency between client and server is already low, the overhead introduced by this additional processing cannot be justified in this context. As shown in Table \ref{tab:overhead_req_mesma_chave_cen_A}, during the first request for a given key, when a cache miss occurs and VNF-Cache must query the server, the average overhead reaches 9.42 ms.

\subsection{Scenario B: Client and VNF-Cache Close, Server Geographically Distant}

The next experiments were performed in a scenario where the database server was geographically distant from both the client and VNF-Cache, which were close to each other. For this purpose, two AWS virtual machines were instantiated to host the database server: one in Ohio, on the east coast of the United States, and another in Tokyo (Japan). The client and VNF-Cache remained on KVM-based virtual machines running on the same physical host in Curitiba, Brazil. Table \ref{tab:cen_B_tempo_capacidade_posicao_AWS} presents the response times obtained for each combination of server location and VNF-Cache capacity. To enable a more detailed analysis of VNF-Cache performance, its storage capacity was varied across 10, 30, 70, and 100 key-value entries.

In the experiments without VNF-Cache, and the server was hosted in Ohio, the average response time for direct client–server requests was approximately 164 ms, with a throughput of around 6 requests per second. When the server was located in Tokyo, the average response time increased to about 292 ms, and the throughput dropped to approximately 3.3 requests per second.

\begin{table}[!htp]
\centering
\caption{Average response times (in milliseconds) for each VNF-Cache capacity and key-value database server placement.}
\label{tab:cen_B_tempo_capacidade_posicao_AWS}
\begin{tabular}{| c | c | c | c | c | c |}
\hline
& \multicolumn{1}{c|}{No}
& \multicolumn{4}{c|}{VNF-Cache capacity (in number of key-value pairs)}\\ \cline{3 - 6} 
Location & VNF-Cache & 10 & 30 & 70 & 100 \\ 
\hline
Ohio & 164 ms & 174,48 ms & 138,51 ms & 64,66 ms & 8,08 ms \\ 
Japan & 292 ms & 303,34 ms & 239,35 ms & 112,66 ms & 11,02 ms \\ 
\hline
\end{tabular} 
\end{table}

The impact of VNF-Cache in this experiment can be considered highly positive. With a cache capacity of 100 key-value sets, the client and VNF-Cache in Curitiba, and the server in Ohio, the average response time dropped to about 8 ms, with a throughput of approximately 118 requests per second. When the server was located in Tokyo, the benefits were even more substantial: the average response time decreased from 292 ms (without VNF-Cache) to 11.02 ms, and the number of requests processed per second increased from 3.3 to an average of 87. Overall, the performance gains reached approximately 95\% for the Ohio server and 96\% for the Tokyo server.

Table \ref{tab:cen_B_tempo_capacidade_posicao_AWS} also highlights the influence of VNF-Cache capacity relative to the size of the database key space. When the cache was limited to only 10 key-value entries, a slight increase in average response time was observed. For instance, with the server in Tokyo and a 10-entry cache located alongside the client, the average response time deteriorated by approximately 3.5\%. With the server in Ohio under the same cache capacity, the degradation reached about 6\%. These results indicate that when the cache is too small to effectively capture the working set of keys, the average response time may become worse than that of direct, cache-free requests.

\begin{figure}[!htb]
\centering
\includegraphics[width=9cm]{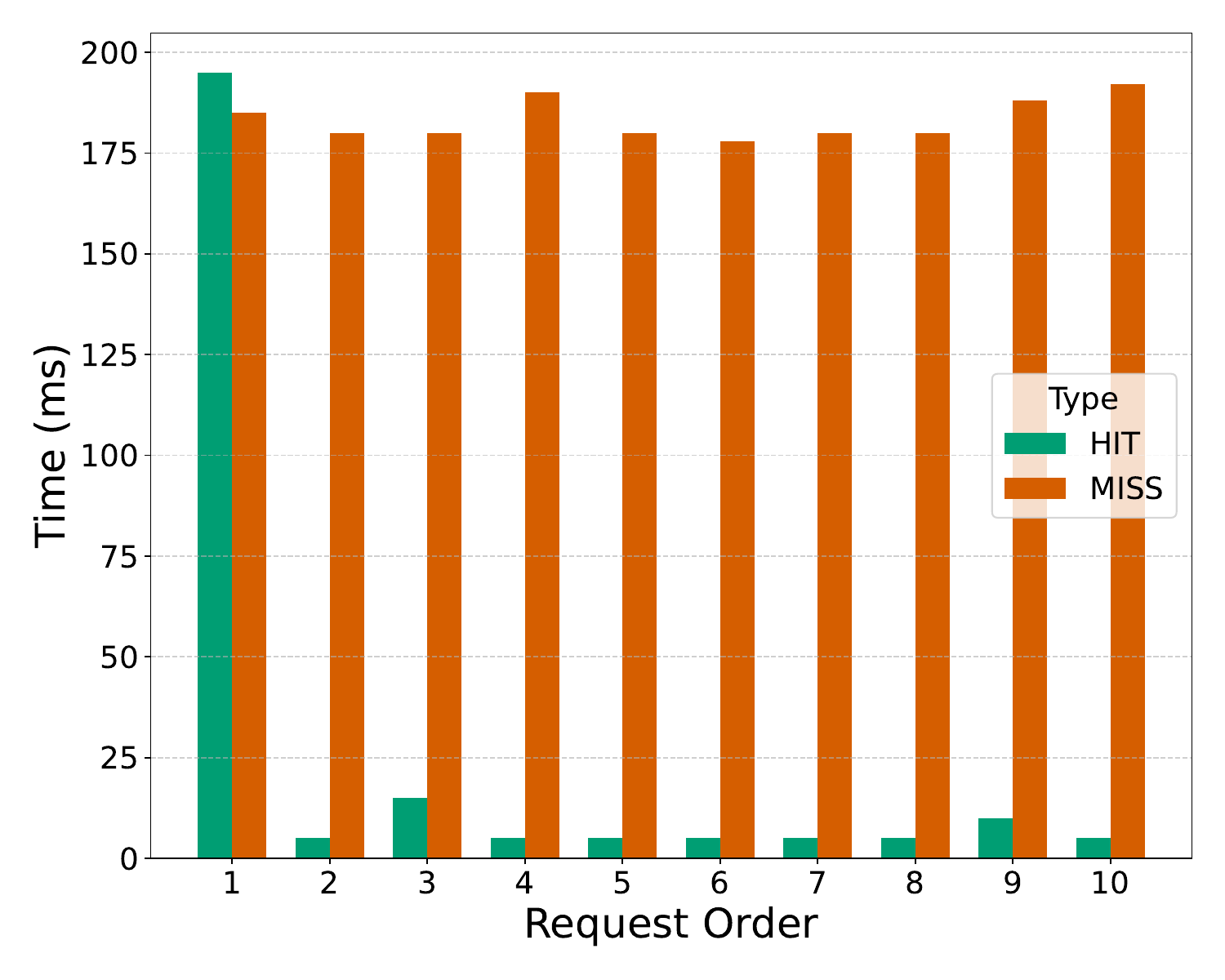}
\caption{Response time for key requests when the query in the VNF-Cache results in a cache miss or cache hit}
\label{fig:miss_hit_reqs}
\end{figure}

Another experiment was conducted to evaluate the performance difference between cache hits and misses, as shown in Figure \ref{fig:miss_hit_reqs}. The figure illustrates the behavior of data requests to the server in Ohio under two conditions: when the VNF-Cache lookup results in a cache miss and when it results in a cache hit. The orange bars represent the response times for the first 10 requests for a key not present in the cache, yielding an average of approximately 180 ms. In contrast, the green bars correspond to the first 10 requests for a key already stored in the cache. In this case, the first request -- responsible for populating the cache -- took nearly 200 ms, but from the second request onward, the average response time dropped to about 6.5 ms. This demonstrates the substantial benefit provided by VNF-Cache: once the key is cached, response time is reduced by roughly 96\%. Moreover, because fewer requests are forwarded to the server, the overall network traffic and server load are also reduced.

\begin{figure}[!htb]
\centering
\includegraphics[width=11cm]{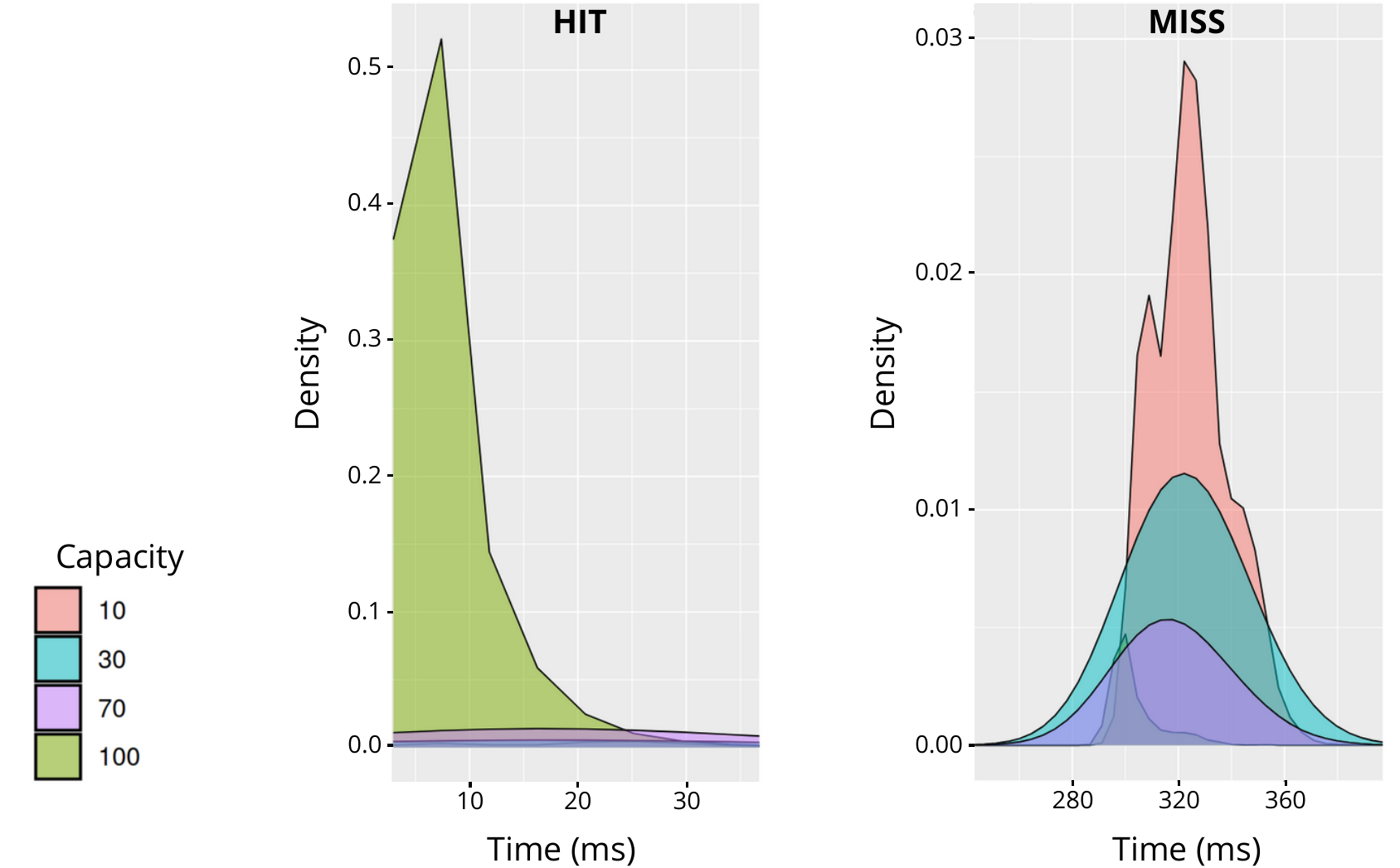}
\caption{Density of response times per VNF-Cache capacity when a cache hit or cache miss occurs. Note the different scales in the graphs.}
\label{fig:cache_miss_hit_density_sizes}
\end{figure}

Figure \ref{fig:cache_miss_hit_density_sizes} presents the density of response times for requests to the server in Tokyo as the VNF-Cache capacity varies. The graph on the left depicts the distribution of requests that resulted in cache hits. When the VNF-Cache is configured with a capacity of 100 key-value sets, the vast majority of hit responses are concentrated between 5 and 15 ms. As the cache capacity decreases, this concentration reduces, as a larger portion of requests result in cache misses and must wait for the server’s response. Another observation is the almost complete absence of hit requests when the cache is limited to 10 entries. In this configuration, most key-value pairs cannot be stored locally, causing their response times to exceed the plotted range.

The graph shown on the right in Figure \ref{fig:cache_miss_hit_density_sizes} shows the cache-miss distribution. The response times for most requests processed by the 10 key-value entries VNF-Cache were between 300 and 350 ms. As the cache capacity increases, the density of cache-miss requests decreases, since a larger share of requests become hits and therefore exhibit significantly lower response times. It is also notable that cache misses are rare when the VNF-Cache is configured with a capacity of 100 entries. In this case, only the first requests for each key are misses, after which subsequent requests benefit from the locally stored values. Consequently, the majority of requests fall outside the cache-miss distribution and appear instead in the cache-hit graph. Overall, the larger the capacity of the VNF-Cache, the greater the concentration of requests with response times shorter than those of direct client–server communication. Conversely, smaller cache capacities lead to a higher density of requests with response times exceeding those of direct access.

\subsection{Scenario C: Client, VNF-Cache, and Server All Geographically Distant from Each Other} 

Experiments of the third scenario were conducted by having the server, VNF-Cache, and client all geographically distant from each other. In this scenario, the client was executed in Curitiba, while the VNF-Cache ran on an AWS virtual machine deployed in São Paulo, Brazil -- rather than being in the same location as the client, as in the previous experiment. The goal was to assess whether a cache positioned in Brazil can benefit local clients accessing servers located in the northern hemisphere. As shown in Table \ref{tab:cen_C_tempo_capacidade_posicao_AWS}, when the VNF-Cache was configured with a capacity of 100 key-value sets, the average request time decreased by approximately 87\% for the server in Ohio and by around 92\% for the server in Tokyo. Likewise, throughput increased to an average of 45 requests per second for Ohio (compared to only 6 with direct access) and to an average of 40 for Tokyo (compared to 3.3 with direct access).

\begin{table}[!htp]
\centering
\caption{Average response times for each capacity of the VNF-Cache in São Paulo and different positioning of the key-value database server.}
\label{tab:cen_C_tempo_capacidade_posicao_AWS}
\begin{tabular}{| c | c | c | c | c | c |}
\hline
& \multicolumn{1}{c|}{No}
& \multicolumn{4}{c|}{VNF-Cache capacity (in number of key-value pairs)}\\ \cline{3 - 6} 
Location & VNF-Cache & 10 & 30 & 70 & 100 \\ 
\hline
Ohio & 164 ms & 134,56 ms & 108,92 ms & 57,88 ms & 21,13 ms \\ 
Japan & 292 ms & 257,08 ms & 204,31 ms & 100,50 ms & 22,13 ms \\ 
\hline
\end{tabular} 
\end{table}

Another observation from Table \ref{tab:cen_C_tempo_capacidade_posicao_AWS} is that VNF-Cache provides benefits even with a small capacity of only 10 key-value sets. In this configuration, the average response time for requests was reduced by approximately 17\% when the server was located in Ohio and by about 12\% when the server was in Tokyo. These improvements are more significant than those observed in Experiment 1, likely due to the performance limitations of the physical machine hosting the KVM virtual machines used for the client and VNF-Cache in that earlier setup.

\begin{figure}[!htb]
\centering
\includegraphics[width=11cm]{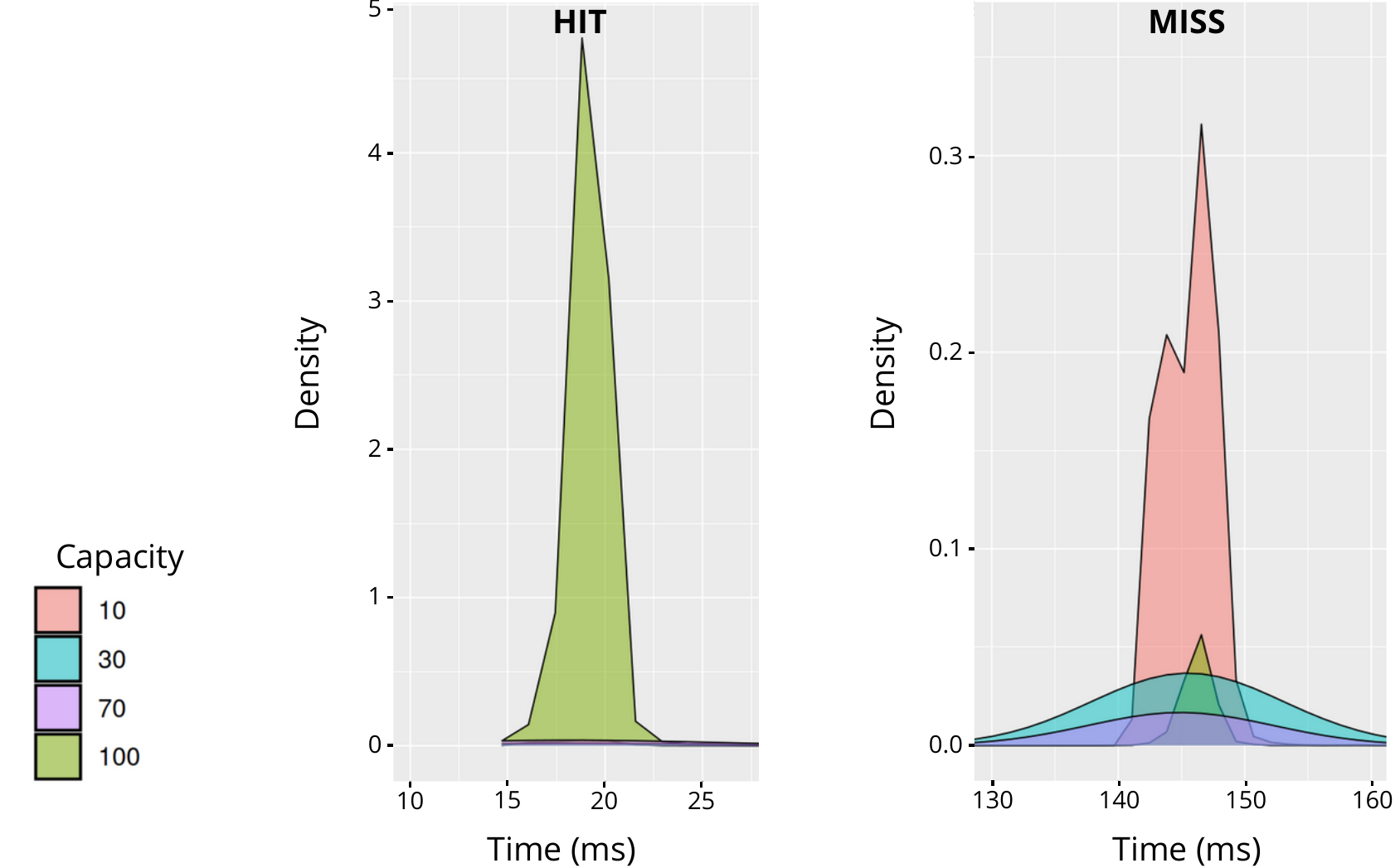}
\caption{Density of response times per VNF-Cache capacity when a cache hit or cache miss occurs (scenario C).}
\label{fig:cache_miss_hit_density_sizes_cachesp}
\end{figure}

Figure \ref{fig:cache_miss_hit_density_sizes_cachesp} presents the distribution of response times for requests to the server in Ohio as the capacity of the VNF-Cache deployed in São Paulo varies. As in Scenario B, when the VNF-Cache is configured with a capacity of 100 key-value pairs, most requests result in cache hits. On the other hand, with a capacity of only 10 key-value pairs, the majority of  requests became cache misses.

\begin{figure}[!htb]
\centering
\includegraphics[width=9cm]{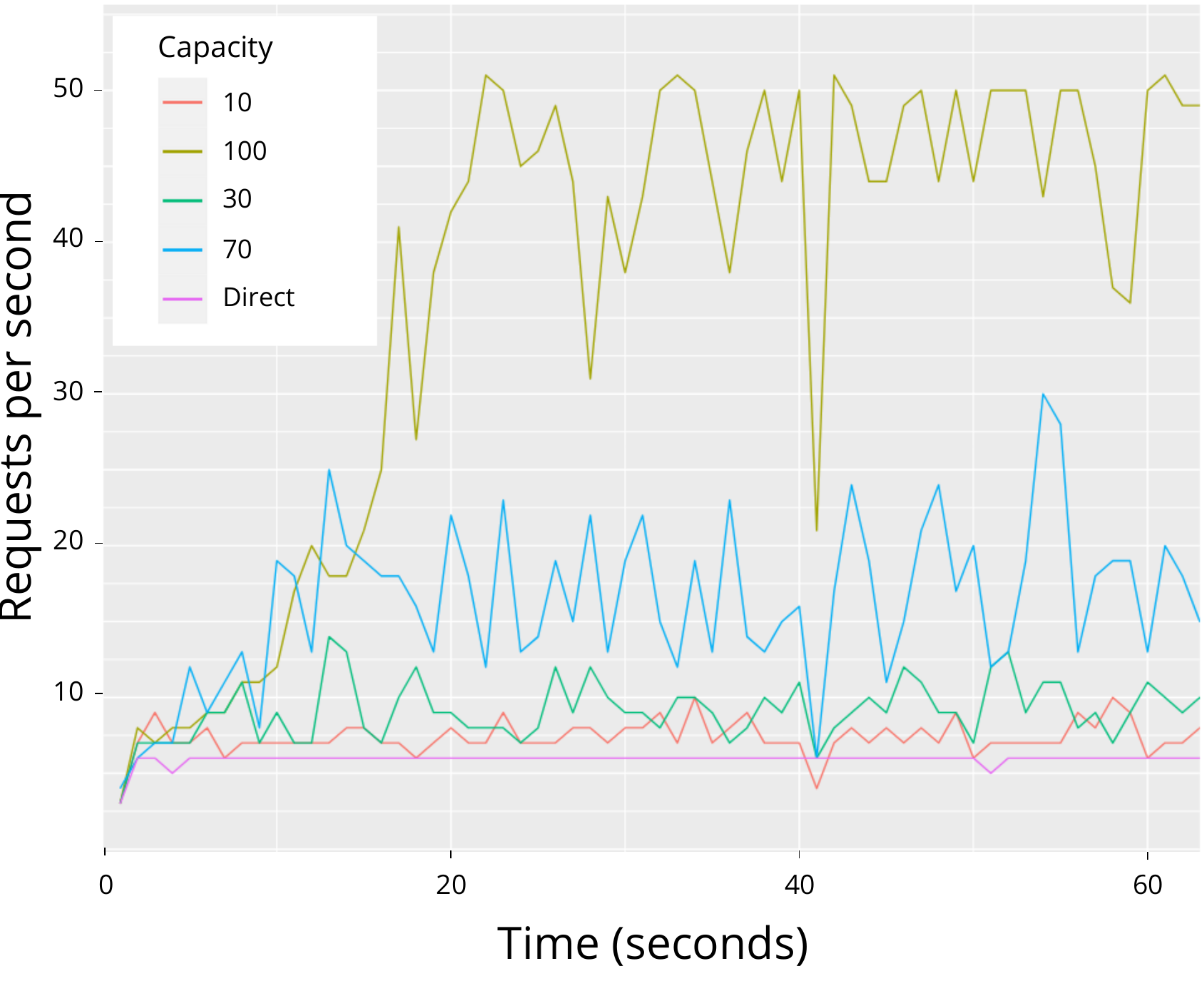}
\caption{Request rate during the first 60 seconds of the experiments for each capacity of the VNF-Cache.}
\label{fig:throughput_60_seconds_sizes}
\end{figure}

Finally, Figure \ref{fig:throughput_60_seconds_sizes} presents the number of requests processed per second during the first 60 seconds of the execution of request batches. The VNF-Cache was running in São Paulo, while the server was located in Ohio.

In the experiment without VNF-Cache, the throughput remained relatively stable, fluctuating between 5 and 6 requests per second. When VNF-Cache was enabled with a capacity of 100 key-value sets, the throughput increased substantially, even up to 50 requests per second, after a warm-up period. In that initial warm-up phase, the low throughput is expected, as the VNF-Cache progressively stores key-value entries locally. Once the cache is sufficiently populated, responses can be returned directly to the client, significantly increasing the throughput and decreasing the response times. The figure also indicates that larger cache capacities lead to higher throughput, reinforcing the impact of cache size on performance.


\subsection{Discussion}

The experiments clearly demonstrate that VNF-Cache effectively reduces response times and increases the number of requests processed per second when accessing remote (geographically distant) key-value databases. Additionally, it is important to note the implicit reduction in network traffic, since many requests and responses no longer need to traverse the full path to the server. Finally, because the VNF-Cache is implemented as a virtual network function, it offers significant deployment flexibility: it can be instantiated and configured at different points in the network rapidly and with minimal operational effort.
\section{Conclusion}
\label{sec-conclusion}

In this work, we proposed the VNF-Cache caching service for key-value databases. VNF-Cache is implemented as a virtual network function. By intercepting and processing network packets exchanged between clients and database servers, VNF-Cache is able to store key-value pairs within the network, effectively bringing data closer to the requesting applications. This mechanism enables direct responses to client requests, reducing response latency, lowering data traffic, and promoting more efficient use of network resources. VNF-Cache was implemented, and the experimental evaluation demonstrated a substantial reduction in the response time for accessing geographically distant key-value database servers. Furthermore, the results indicate a significant increase in the number of requests processed per second, highlighting the effectiveness of the proposed approach.

Future work includes the implementation of more advanced cache population and replacement policies, enabling finer control over which data items are stored and retained. Another planned extension is to broaden the range of supported back-end systems by enabling the caching of data from additional databases and heterogeneous key-value stores -- such as Redis or Amazon DynamoDB -- potentially allowing VNF-Cache to operate across multiple data sources simultaneously. A current limitation of the prototype is the absence of mechanisms to ensure secure communication between clients and servers. Addressing this is essential to guarantee the safe deployment of VNF-Cache in practical environments. Finally, further investigation is required into more robust data structures capable of efficiently storing key-value sets while supporting concurrent operations, which would enhance both performance and scalability.
\section*{Acknowledgments}
This work has been partially supported by the Coordination for the Improvement of Higher Education Personnel (CAPES) - Program of Academic Excellence (PROEX), Funding Code 001; and the Brazilian National Council for Scientific and Technological Development (CNPq) - grant 305108/2025-5.

\bibliographystyle{IEEEtran}
\bibliography{refs-arXiv-VNFCache}

\end{document}